\documentclass[12pt]{article}
\usepackage[top=1.5in,bottom=1.5in,left=1.25in,right=1.25in]{geometry}
\usepackage{graphicx}
\usepackage{amsmath}
\usepackage{bbm}
\usepackage{amssymb}
\usepackage{amsthm}
\usepackage{amsfonts}
\usepackage{wrapfig}
\usepackage{multicol}
\usepackage{float}
\usepackage{pdfpages}
\usepackage[colorlinks=true, urlcolor=blue, linkcolor=blue,citecolor=blue]{hyperref}
\usepackage{natbib}
\usepackage[utf8]{inputenc}
\usepackage{mathrsfs}

\usepackage{color}

\newtheorem*{definition}{Definition}
\theoremstyle{remark}
\newtheorem{example}{Example}
\theoremstyle{remark}

 \usepackage{booktabs}

\DeclareMathOperator*{\argmin}{arg\,min}

\makeatletter
\@addtoreset{section}{part}
\def\@part[#1]#2{%
    \ifnum \c@secnumdepth >\m@ne
      \refstepcounter{part}%
      \addcontentsline{toc}{part}{\thepart\hspace{1em}#1}%
    \else
      \addcontentsline{toc}{part}{#1}%
    \fi
    {\parindent \z@ \raggedright
     \interlinepenalty \@M
     \normalfont\centering
     \ifnum \c@secnumdepth >\m@ne
       \LARGE\bfseries \partname\nobreakspace\thepart
       \par\nobreak
     \fi
     \huge \bfseries #2%
     \markboth{}{}\par}%
    \nobreak
    \vskip 3ex
    \@afterheading}
\renewcommand\partname{Appendix}
\makeatother

\newcommand{\omt}[1]{}
\title{Measuring the Completeness of Theories\thanks{This is an updated version of ``The Theory is Predictive, but is it Complete?" We thank  Alberto Abadie, Amy Finkelstein, and Johan Ugander for helpful comments. We are also grateful to Adrian Bruhin, Helga Fehr-Duda, Thomas Epper, Kevin Leyton-Brown, and James Wright for sharing data with us.}}
\author{Drew Fudenberg\thanks{Department of Economics, MIT} \quad Jon Kleinberg\thanks{Department of Computer Science, Cornell University} \quad Annie Liang\thanks{Department of Economics, University of Pennsylvania} \quad Sendhil Mullainathan\thanks{Department of Economics, University of Chicago}}
 
\begin{document}

\maketitle

\begin{abstract}

We use machine learning to provide a tractable measure of  the amount of predictable variation in the data that a theory captures, which we call its  ``completeness."  We apply this measure to three problems: assigning certain equivalents to lotteries, initial play in games, and human generation of random sequences. We discover considerable variation in the completeness of existing models, which sheds light on whether to focus on developing better models with the same features or instead to look for new features that will improve predictions.  We  also  illustrate  how  and  why  completeness varies with the experiments considered, which  highlights the role played in choosing which experiments to run. 

\end{abstract}

\def\citeasnoun{\cite}

Suppose we have a theory of the labor market
that says that a person's wages depend on 
their knowledge and capabilities. We can test this theory by looking at whether more education indeed predicts higher wages in labor data. If it does, this would provide evidence in support of the theory, but it would not tell us whether an alternative theory might be even more predictive.
The question of whether there are  more predictive theories, and if so how much
more predictive they might be, raises the issue of {\em completeness}: How close is the performance
of a given theory to the best performance that is achievable in the domain?
In other words, how much of the
predictable variation in the data is captured by the theory?

We cannot gauge completeness of a theory solely through the level of its predictive accuracy because there is intrinsic noise in the outcome.  For example,  an accuracy of 55\% is strikingly successful for predicting a (discretized) stock movement based on past returns, but extremely weak for predicting the (discretized) movement of Earth based on the masses and positions of the sun and all of the other planets.  These two problems differ in the best achievable prediction performance they permit, and so the same quantitative level of predictive accuracy must be interpreted differently in the two domains.

One way to view the contrast between these two problem domains is as follows. In each case, an instance $i$ of the prediction problem 
consists of a vector $x_i$ of measured features or covariates, and a hidden outcome $y_i$ 
that must be predicted.
In the case of astronomical bodies, we believe that the measured
features are sufficient to make highly accurate 
predictions over short time scales.
In the case of stock prices, the measured features\textemdash
past prices and returns\textemdash are only a small fraction of the 
information that we believe may be relevant to future prices. 
Thus, the variation in stock movements {\em conditioned
on the features we know} is large, while planetary motions are well-predicted by known features.

The point then is that prediction error represents
a composite of two things: first, the opportunity for a more predictive model;
and second, intrinsic noise in the problem due to the limitations
of the feature set.  If we want to understand how much room 
there is for improving the predictive performance of existing theories within a given domain\textemdash
holding constant the set of features that we know how to measure\textemdash
we need a way to separate these two effects.

We propose that a good way to distinguish between these sources of prediction error is to  compare the performance of the existing models with the \emph{best achievable level} of prediction for our feature set,  as computed by a  \emph{Table Lookup} algorithm. This algorithm  finds the best prediction for each feature vector, assuming that the distribution of training instances approximates the actual relationship between the observable features and the outcome. With an ideal (i.e. infinite) data set, Table Lookup minimizes the expected out-of-sample error, but Table Lookup can be quite imperfect when data is sparse.  Appendix \ref{app:TLerrBound} provides remarks that justify the use of the performance of Table Lookup as an approximation to the best achievable accuracy in our applications.

We illustrate the usefulness of Table Lookup by applying it to three different problem domains: the evaluation of risk, initial play in games, and human perception of randomness. These are all  are important topics in economics, with a long line of established models. We use our benchmark to evaluate the completeness of leading models from each domain.  Interestingly, we find that the best model we use for the perception of randomness is only $24\%$ complete, while Cumulative Prospect Theory is $95\%$ complete despite having a mean-squared prediction error of 67.38. This, and the subsequent observations we make in Sections \ref{sec:risk}-\ref{sec:coins}, are informative about the problem domains  and the status of their associated models.


Our main contribution, however, is methodological: since most economic behaviors cannot be perfectly predicted given  the available features, the predictive accuracies of our models are difficult to interpret on their own. Understanding not just how well existing models predict, but also how complete they are, is important for guiding the development of the theory. In the three applications in this paper, we show how  these benchmarks can  be constructed, and that they reveal non-obvious insights into the performance of our existing models.  We also illustrate how and why our  our completeness measure varies with the experiments considered, for example with the choice of lotteries used to evaluate risk preferences. This dependence highlights the key role played in choosing which experiments to run.  

The Table Lookup benchmark is applicable to domains beyond the three that we describe here, but not in all of them, and we discuss various limitations to its applicability and interpretation throughout the paper. First, as we explain in Section \ref{subsec:finite}, Table Lookup approximates the best achievable predictive accuracy only when the data set is large relative to the number of unique feature vectors. This requires either that features are discrete-valued (as in our application in Section \ref{sec:coins}), or that the available data involves observations from a finite number of unique instances (as in our two applications in Sections \ref{sec:risk} and \ref{sec:games}). Although Table Lookup is not feasible for all problems, the range of applications in the paper suggest that Table Lookup is more effective than one might initially suspect.

Second, our completeness measure depends on a specified set of features, and is evaluated on a given data set. If we change either the underlying feature set or the data, we would expect the measurement of completeness to also change, as we discuss in Section \ref{sec:featureset}. The dependence of completeness on what data set we use is important to keep in mind.  Moreover, as we show in Sections \ref{sec:games}, the way the completeness of a model varies across data sets can shed light on the domains in which the model performs well or performs poorly.

\section{Problem and Approach} \label{sec:problem}

\subsection{Prediction Problems}

In a prediction problem, there is an \emph{outcome} $Y$ whose realization is of interest, and \emph{features} $X_1, \dots, X_N$ that are statistically related to the outcome. The goal is to predict the outcome given the observed features. Some examples include predicting  an individual's future wage based on childhood covariates (city of birth, family income, quality of education, etc.), or predicting a criminal defendant's flight risk based on their past record and properties of the crime \citep{bail}. We focus here on three prediction problems that emerge from experimental economics:

\begin{example}[Risk Preferences] Can we predict the valuations that people assign to various money lotteries?

\end{example}

\begin{example}[Predicting Play in Games]  Can we predict how people  will play the first time they encounter a given simultaneous-move game? 
\end{example}

\begin{example}[Human Generation of Random Sequences] Given a target random process\textemdash for example, a Bernoulli random sequence\textemdash can we predict the errors that a human makes while mimicking this process?
\end{example}

Formally, suppose that the observable features belong to some space $\mathcal{X} = \mathcal{X}_1 \times \cdots \times \mathcal{X}_N$ and the outcome belongs to $\mathcal{Y}$. A  map $f: \mathcal{X} \rightarrow \mathcal{Y}$ from features to outcomes is  a \emph{(point) prediction rule}.\footnote{Note that a  prediction of a probability distribution over $\mathcal{Y}$ can be cast as the prediction of a point in the space $Y'=\Delta(\mathcal{Y})$ of distributions on $\mathcal{Y}$.}  Many economic models can be described as a parametric family of prediction rules $\mathcal{F}=(f_\theta)_{\theta \in \Theta}$. For example, if our model class $\mathcal{F}$ imposes a linear relationship between the outcome and a set of features, then the parameter $\theta$ would define a vector of weights applied to each of the features. In the application we study in Section \ref{sec:risk}, the expected utility class $\mathcal{F}$ describes a family of utility functions $u(z)=z^\theta$ over dollar amounts, and the parameter $\theta$ reflects the degree of risk aversion.

\subsection{Accuracy and Completeness} \label{subsec:AccComplete}

We suppose that our prediction problem comes with a a \emph{loss function}, $\ell: \mathcal{Y} \times \mathcal{Y} \rightarrow \mathbb{R}$, where $\ell(y',y)$ is the error assigned to prediction of $y'$ when the realized outcome is $y$. The commonly used loss functions \emph{mean-squared error} and  \emph{classification loss}  correspond to $\ell(y',y)=(y'-y)^2$ and $\ell(y',y)=\mathbbm{1}(y'\neq y)$ respectively.

\begin{definition}The \textbf{expected error} (or \textbf{risk}) of prediction rule $f$ on a new observation $(x,y)$ generated according to the joint distribution $P$ is\footnote{Different loss functions are typically used when predicting  distributions, see e.g. \citet{Gneiting2007}.}
\[\mathcal{E_{P}}(f) = \mathbb{E}_{P}[\ell(f(x),y)].\]
\end{definition}
 The prediction rule in the class $\mathcal{F}$ that minimizes the expected prediction error is the one associated with the parameter value
 \[\theta^\ast_P= \argmin_{\theta \in \Theta} \mathcal{E_{P}}(f_\theta).\]
The expected error of this ``best" rule in $\mathcal{F}$ is $\mathcal{E_{P}}(f_{{\theta^\ast_P}})$.

In Section \ref{cross-val}, we discuss how to estimate $\mathcal{E_{P}}(f_{{\theta^\ast_P}})$  on finite data; here we discuss how to interpret it.  To understand a model's error,  it is helpful to distinguish between two very different error sources.

First, if the the conditional distribution $Y \mid X$ is not degenerate, then even the ideal prediction rule 
\[f_{P}^\ast(x) = \argmin_{y' \in \mathcal{Y}} \mathbb{E}_P \left[ \ell(y',y) \mid x\right]\]
does not predict perfectly.

\begin{definition} The \textbf{irreducible error} in the prediction problem is the expected error 
\begin{equation}
    \label{disp:IrrErr}
\mathcal{E_{P}}(f^\ast_P)=\mathbb{E}_P\left[\ell(f^\ast_P(x),y)\right]
\end{equation}
of the  ideal rule on a new test observation.
\end{definition}
\noindent The irreducible error is an upper bound on how well we can predict $Y$ using the features $X$. 

A different source of prediction error is the specification of which  prediction rules $f: \mathcal{X} \rightarrow \mathcal{Y}$ are in the class $\mathcal{F}$.  Typically the best possible model will not be an element of  $\mathcal{F}$---that is, most sets of models are at least slightly misspecified. 
If  $\mathcal{F}$ leaves out an important regularity in the data, there may be exist models outside of $\mathcal{F}$ that give much better predictions on this domain.\footnote{On the other hand, expanding the model class risks overfitting, so more parsimonious model classes can lead to more accurate predictions when data is scarce  \citep{Hastie}. As we discuss in Sections \ref{sec:TL} and \ref{sec:related}, all  of the data sets we consider here are large relative to the number of features.}

These two sources of prediction error have very different implications for how to improve prediction in the domain. If the achieved performance of the model is substantially lower than the best feasible performance, then it may be possible to achieve large improvements \emph{without} seeking additional inputs, for example by identifying new regularities in behavior. On the other hand, if the achieved prediction error is close to the \emph{best achievable level} of prediction for our feature set, then only marginal gains are feasible from identification of new structure. This encourages consideration of prediction rules $f: \mathcal{X}'\rightarrow \mathcal{Y}$ based on some larger feature space $\mathcal{X}'$. 

We propose the ratio of reduction in prediction error achieved by the model, compared to the \emph{achievable} reduction, as a measure of how close the model comes to the best achievable performance. We call this ratio the the model's \emph{completeness}. To operationalize this measure, let $f_{\text{naive}}: \mathcal{X}\rightarrow \mathcal{Y}$ be a naive rule suited to the prediction problem; this rule\textemdash such as ``predict uniformly at random"\textemdash is meant to represent a lower bound on how bad predictions can be.

\begin{definition} The \bf{completeness} for the model class $\mathcal{F}$ is
\begin{equation}
    \label{exp:Completeness}
\frac{\mathcal{E_{P}}(f_{\text{naive}})-\mathcal{E_{P}}(f_{\theta^\ast_P})}{\mathcal{E_{P}}(f_{\text{naive}})-\mathcal{E_{P}}(f_{P}^\ast)}.
\end{equation}
\end{definition}
Note that the completeness measure depends on the underlying distribution $P$. We  expect the conditional distribution $P(y \mid x)$ to be a fixed distribution describing the true dependence of the outcome on the features, but the marginal distribution over the feature space $\mathcal{X}$ is frequently a choice variable of the analyst\textemdash e.g. which lotteries or games to run in an experiment. As we show in Section \ref{sec:games}, when we change this marginal distribution, we obtain different measures of completeness for the same model. Ideally, we would like the chosen distribution over features to be the one that is most economically relevant, but in practice we may not know what that is.

\subsection{Evaluating Completeness on Finite Data} \label{subsec:finite}

Neither the true joint distribution $P$ over features and outcomes nor the derived  quantities $\mathcal{E_{P}}(f_{\text{naive}})$, $\mathcal{E_{P}}(f_{\theta^\ast_P})$, and $\mathcal{E_{P}}(f_{P}^\ast)$ are  directly observable, but they  can be estimated from data. We describe below an approach (tenfold cross-validation) that is standard for estimating expected prediction errors, and describe an algorithm\textemdash \emph{Table Lookup}\textemdash for approximating the ideal prediction rule $f^\ast_P$.

\subsubsection{Cross-Validated Prediction Errors}\label{cross-val}

To evaluate the predictive accuracy of a model class $\mathcal{F}$ on a finite data set, we first choose between prediction rules in $\mathcal{F}$ based on how well they predict a sample of training observations. Then we evaluate the trained rule on a new set of test observations.

Formally,
 for any integer $n$ let   $\mathcal{Z}^n =(\mathcal{X}\times \mathcal{Y})^n $ be the space corresponding to $n$ observations of $(x,y)$, and suppose the analyst has access to a data set $Z \in \mathcal{Z}^n$. Using the procedure of $K$-fold cross-validation, this data is randomly split into $K$ equally-sized disjoint subsets $Z_1, \dots, Z_{K}$. In each iteration $1\leq i \leq K$ of the procedure, the subset $Z_i$ is identified as the \emph{test data} and the remaining subsets are used as \emph{training data}. The $i$-th parameter estimate is the one that minimizes average loss when predicting the $i$-th training set:
 \[\theta^\ast_i= \argmin_{\theta \in \Theta} \frac{1}{n} \sum_{(x,y) \in \cup_{j\neq i} Z_j} \ell(f_{\theta} (x), y).\]
(The naive model prediction rule does not depend on the training data, and is always $f_{\text{naive}}$.) The out-of-sample error of the estimated $f_{\theta^*_i}$ on the test set $Z_i$ is
 \begin{equation}
     \label{eq:CVfold}
\text{CVErr}_i = \frac1m \sum_{(x,y)\in Z_i} \ell(f_{\theta^*_i} (\mathbf{x}),y).
 \end{equation}
If the data in $Z$ are drawn i.i.d. from $P$, the average out-of-sample error 
\begin{equation}
    \label{eq:CV}
\text{CVErr(Z)}=\frac1K \sum_{i=1}^K \text{CVErr}_i
\end{equation}
is a consistent estimator for the expected error $\mathcal{E_{P}}(f_{\theta^{\ast}_P})$. The display in (\ref{eq:CV}) is known as the \emph{$K$-fold cross-validated prediction error}. In the main text, we will more simply refer to it as the \emph{prediction error} of the model class $\mathcal{F}$, understanding that it is a finite-data estimate.

Below we write $CV_{\text{naive}}(Z)$ for the cross-validated prediction error for the naive rule $f_{\text{naive}}$ and $CV_{\mathcal{F}}(Z)$ for the cross-validated prediction error for the model class $\mathcal{F}$. These are respectively our estimates for $\mathcal{E}_P(f_{\text{naive}})$ and $\mathcal{E}(f_{\theta^\ast_P})$.  

\subsubsection{Table Lookup Benchmark} \label{sec:TL}


To estimate the expected error of the ideal rule $\mathcal{E_{P}}(f_{P}^*)$, we apply a Table Lookup algorithm to each iteration $i$ of cross-validation: Formally,  let
\[f^{TL}_i=\argmin_{f \in \mathcal{X}^\mathcal{Y}} \frac{1}{n} \sum_{(x,y) \in \cup_{j\neq i} Z_j} \ell(f_{\theta} (x), y)\]
be the function that minimizes prediction error on the training data, where we search across the complete (unrestricted) class of mappings from $\mathcal{X}$ to $\mathcal{Y}$. Then define the cross-validated Table Lookup error as in (\ref{eq:CVfold}) and (\ref{eq:CV}). This measure, which we will denote $CV_{TL}$, is  a consistent estimator for the irreducible error $\mathcal{E}_P(f^\ast_P)$. How good of an approximation it is depends on a comparison between the size of the data $n$ and the ``effective" size of the feature set $\mathcal{X}$, by which we mean the number of unique feature vectors $x$ that appear in the data.\footnote{Table Lookup predicts well when we have a large number of observations for each unique feature vector $x \in \mathcal{X}$. This requires either that the feature space $\mathcal{X}$ is finite (as in our application in Section \ref{sec:coins}, where $\mathcal{X}=\{0,1\}^7$), or that the data-generating measure $P$ has finite support over $\mathcal{X}$ (as in our two applications in Sections \ref{sec:risk} and \ref{sec:games}). In some settings with a continuum of possible features there may be very few observations for a given feature vector. In these cases, we cannot directly use Table Lookup to approximate the ideal performance, and should instead use approaches that make assumptions on how outcomes are related at ``nearby" features, e.g. kernel regression.} 

One way to evaluate the accuracy of $CV_{TL}$ is to look at the \emph{standard error of the cross-validated prediction errors}, which is
\[\sqrt{\frac{1}{K}\text{Var}(\mbox{CVErr}_1,\dots,\mbox{CVErr}_K)}.\]
 We report these standard errors for each of our applications and model classes. It turns out that for each of the applications we look at, and we suspect for other data sets as well, the Table Lookup standard errors are relatively small. (See
Appendix \ref{app:CVSE} for more detail.) As another test, we compare the performance of Table Lookup with a different machine learning algorithm that is better suited to smaller data sets (bagged decision trees), and find that Table Lookup's performance is comparable but better for all of our applications (see Appendix \ref{app:RF}). These analyses suggest that the Table Lookup performance is indeed a reasonable approximation for the best achievable performance in each of our applications.

In place of the ideal completeness measure described in (\ref{exp:Completeness}), we compute the following ratio from our data:
\[\frac{\mbox{CV}_{\text{naive}} - \mbox{CV}_{\mathcal{F}}}{\mbox{CV}_{\text{naive}} - \mbox{CV}_{TL}}.\]
This is the ratio of reduction in cross-validated prediction error achieved by the model (relative to the naive baseline) compared to the reduction achieved by Table Lookup (again relative to the naive baseline).

\subsection{Relationship to Literature} \label{sec:related}

Irreducible error is an old concept in statistics and machine learning, and a large amount of work has focused on further decomposing this error into \emph{bias} (reflecting error due to the specification of the model class) and \emph{variance} (reflecting sensitivity of the estimated rule to the randomness in the training data). Depending on the quantity of data available to the analyst, it may be preferable to trade off bias for variance or vice versa.\footnote{For example, given small quantities of data, we may prefer to work with models that have fewer free parameters, leading to higher bias but potentially substantially lower variance.} This paper abstracts from these concerns, as well as the related concern of overfitting. We work exclusively with data sets where the quantity of data is large enough that the most predictive model is approximately the most complex one, i.e. Table Lookup (see Appendix \ref{app:TLerrBound}). 

  A related literature compares the performance of specific machine learning algorithms to that of existing economic models.  These algorithms are themselves potentially incomplete relative to the best achievable level, and thus provide a \emph{lower bound} for the best achievable level, where the degree to which they are incomplete is a priori unknown. The closest of these papers to our work is \citet{Naecker}, which studies choices under uncertainty and under ambiguity, and constructs a benchmark based on regularized regression algorithms.\footnote{In addition, \citet{Plonskyetal2017}, \citet{Notietal}, and \citet{Plonskyetal} develops  algorithmic models for predicting choice, \citet{CamererNaveSmith} uses machine learning to predict disagreements in bargaining, and \citet{BodohCreed} uses random forests to predict pricing variation. The improvements achieved by the algorithms are sometimes modest, perhaps due to intrinsic noise,  as \citet{Bourginetal} point out. We show how this noise can be quantified.}

\citet{ErevRoth2007} define a a model's \emph{equivalent number of observations} as  the number $n$ of prior observations such that the mean of  a data set of $n$ random observations has the same prediction error as the model. We expect that models with larger numbers of equivalent observations will be more complete by our metric.
 
Finally, an alternative measure of a model's performance is the  proportion of the variance in the outcome that it explains, that is the model's $R^2$.  This measure is not well suited to the question of the model's completeness, because the \emph{best achievable} $R^2$ cannot be directly inferred from the $R^2$ of any existing model.\footnote{We could, however, develop a notion of completeness based on comparing the achieved $R^2$ with the best achievable $R^2$, analogous to what we do here.}

\section{Three Applications}

\subsection{Domain \#1: Assigning Certain Equivalents to Lotteries} \label{sec:risk}

\paragraph{Background and Data.} An important question in economics is how individuals evaluate risk. In addition to the Expected Utility models \citep{vNM,Savage,Samuelson}, one of the most influential models of decision-making under risk in the last few decades has been Cumulative Prospect Theory \citep{CPT}. This model provides a flexible family of risk preferences that accommodates certain behavioral anomalies, including reference-dependent preferences and nonlinear probability weighting.

A standard experimental paradigm for eliciting risk preferences, and thus for evaluating these models, is to ask subjects to report \emph{certainty equivalents} for lotteries\textemdash i.e. the lowest certain payment that the individual would prefer over the lottery. We consider a data set from \citet{Bruhin}, which includes 8906 certainty equivalents elicited from 179 subjects, all of whom were students at the University of Zurich or the Swiss Federal Institute of Technology Zurich. Subjects reported certainty equivalents for the same 50 two-outcome lotteries, half over positive outcomes (e.g. gains) and half over negative outcomes (e.g. losses).

\paragraph{Prediction Task and Models.} 

In this data set, the outcomes are the reported certainty equivalents for a given lottery, and the features are the lottery's two possible monetary prizes $z^1$ and $z^2$, and the probability $p$ of the first prize. A prediction rule is any function that maps the tuple $(z^1,z^2,p)$ into a prediction for the certainty equivalent, i.e. a function $f: \mathbb{R}\times \mathbb{R}\times[0,1]\rightarrow \mathbb{R}$.

We evaluate two prediction rules that are based on established models from the literature. Our \emph{Expected Utility} (EU) rule sets the agent's utility function to be $u(z)=z^\alpha$, where $\alpha$ is a free parameter that we train. The predicted certainty equivalent is $pz_1^\alpha + (1-p) z_2^\alpha$.

Second, our \emph{Cumulative Prospect Theory} (CPT) rule predicts $w(p)v(z_1) + (1-w(p)) v(z_2)$ for each lottery, where $w$ is a probability weighting function and $v$ is a value function. We follow the literature (see e.g. \citet{Bruhin}) in assuming the functional forms:
\begin{equation} \label{eq:functionalform}
v(z)=\left\{ \begin{array}{cc}
z^\alpha & \mbox{ if } z>0 \\
-(-z^\beta) & \mbox{ if } z \leq 0
\end{array}\right. \quad \quad \quad w(p)= \frac{\delta p^\gamma}{\delta p^\gamma + (1-p)^\gamma}.
\end{equation}
This model has four free parameters $\alpha,\beta,\delta,\gamma \in \mathbb{R}_+$. 

Finally, as a naive benchmark, we predict the \emph{expected value} of the lottery, which is $p_1z^1 + (1-p_1) z^2$.\footnote{This naive benchmark is arguably less naive than the naive benchmarks we use for the other prediction problems. Replacing our naive benchmark with, for example, an unconditional mean, would result in even higher completeness for CPT than we already find in Table \ref{tab:resultsCE}.}

\paragraph{Performance Metric.} \label{sec:CV}

For a given test set of $n$ observations $\{(z^1_i, z^2_i,p_i; y_i)\}_{i=1}^n$\textemdash where $(z^1_i, z^2_i,p_i)$ is the lottery shown in observation $i$, and $y_i$ is the reported certainty equivalent\textemdash we evaluate the \emph{prediction error} of prediction rule $f$ using \begin{equation*}
\frac1n \sum_{i=1}^n \left( f(z^1_i,z^2_i,p_i) - y_i)\right)^2.
\end{equation*}
This loss function, \emph{mean-squared error}, penalizes quadratic distance from the predicted and actual response, and is minimized when $f(z^1,z^2,p)$ is the mean response for lottery $(z^1,z^2,p)$. 

To conduct out-of-sample tests of the models described above, we follow the standard approach of tenfold cross-validation described in Section \ref{cross-val}, estimating the free parameters of the model on training data and evaluating how well the estimated model predicts choices in a test set.

\paragraph{Results.} The following table reveals that both models are predictive, improving upon the Expected Value benchmark:\footnote{The parameter estimate for EU is $\alpha=0.98$, and the parameter estimates for CPT are $\alpha=1.024, \beta=0.975, \delta = 0.5,$ and $\gamma=0.525$.} 

\begin{table}[H]
\begin{center}
\begin{tabular}{lc}
\toprule
& Error \\
\midrule
Naive Benchmark & 103.81\\
& (4.00) \\
Expected Utility & 99.67  \\
& (4.50) \\
CPT & 67.38  \\
& (4.49) \\
\bottomrule
\end{tabular} 
\end{center}
\caption{Both models are predictive.}
\end{table}
The improvement of CPT over the naive benchmark is larger than that of Expected Utility, but the CPT performance is substantially worse than perfect prediction. It is not surprising that these models do not achieve perfect prediction, as we expect different subjects to report different certainty equivalents for the same lottery, and thus a model that provides the same prediction for each $(z^1,z^2,p)$ input cannot possibly predict every reported certainty equivalent.

But another source for prediction error is the functional form assumptions that we made in (\ref{eq:functionalform}). Could a different (potentially more complex) specification for the value function or probability weighting function lead to large gains in prediction? Moreover, might there be other features of risk evaluation, yet unmodelled, which lead to even larger improvements in prediction?

To separate these sources of error, we need to understand how the CPT performance compares to the best \emph{achievable} performance for this data. For this evaluation, we construct an ideal benchmark using a Table Lookup procedure. The lookup table's rows correspond to the 50 unique lotteries in our data, and the predicted certainty equivalent for each lottery is the mean response for that lottery in the training data. Given sufficiently many reports for each lottery, the lookup table prediction approximates the actual mean responses in the test data, and its error approximates the best possible error that is achievable by any prediction rule that takes $(z^1,z^2,p)$ as its input. We report this benchmark below in Table \ref{tab:resultsCE}:

\begin{table}[H]
\begin{center}
\begin{tabular}{lcc}
\toprule
 & Error & Completeness\\
\midrule
Naive Benchmark & 103.81 & 0\%\\
& (4.00) \\
Expected Utility & 99.67 & 11\% \\
& (4.50) \\
CPT & 67.38 & 95\% \\
& (4.49) \\
Table Lookup & 65.58 & 100\%\\
& (3.00) \\
\bottomrule
\end{tabular} 
\end{center}
\caption{CPT is nearly complete for prediction of our data.} \label{tab:resultsCE}
\end{table}

The Table Lookup benchmark shows that no prediction rule based on $(z^1,z_2,p)$ can improve more than slightly over CPT on this data, because  CPT obtains 95\% of the feasible improvement in prediction.\footnote{From this data it is hard to know whether the high completeness of  CPT (in the specified functional form)  comes from its good match to actual behavior or because it is flexible enough to mimic Table Lookup on many data sets. We leave exploration of this question to future work.} This tells us that to make substantially better predictions, we would need to expand the set of variables on which the model depends. For example, as we discuss in Section \ref{tab:subsample}, we could group subjects using auxiliary data such as their evaluations of other lotteries or response times, and make separate predictions for each group.

We note that our completeness measure does not imply that \emph{in general} CPT is a nearly-complete model for predicting certainty equivalents, since the completeness measure we obtain is determined from a specific data set, and thus its generalizability depends on the extent to which that data is representative. Indeed, the data from \citet{Bruhin} has certain special features; for example, all lotteries in the data are over two possible outcomes. It would be an interesting exercise to evaluate the completeness of CPT using observations on lotteries with more complex supports.

\subsection{Domain \#2: Initial Play in Games} \label{sec:games}

\paragraph{Background and Data.} In many game theory experiments, equilibrium analysis has been shown to be a poor predictor of the choices that people make when they encounter a new game. This has led to models of initial play that depart from equilibrium theory, for example the  level-$k$ models of \citet{StahlWilson94} and \citet{Nagel},
the Poisson Cognitive Hierarchy model \citep{CamererHoChong04}, and the related models surveyed in \citet{CrawfordSurvey}. These models represent improvements over the equilibrium predictions, but we do not know how substantial these improvements are. Are there important regularities in play that have not yet been modeled?

To study this question, we use a data set from \citet{FudenbergLiang} consisting of 23,137 total observations of initial play from 486 $3 \times 3$ matrix games.\footnote{This data is an aggregate of three data sets: the first is a meta data set of play in 86 games, collected from six experimental game theory papers by Kevin Leyton-Brown and James Wright, see \citet{LeytonBrownWright}; the second is a data set of play in 200 games with randomly generated payoffs, which were gathered on MTurk for \citet{FudenbergLiang}; the final is a data set of play in 200 games that were ``algorithmically designed" for a certain model (level 1) to perform poorly, again from \citet{FudenbergLiang}.}$^,$\footnote{There was no learning in these experiments\textemdash subjects were randomly matched to opponents, were not informed of their partners' play, and did not learn their own payoffs until the end of the session.}   As in the previous section, we pool observations across all of the subjects and games. 

\paragraph{Prediction Task, Performance Metric, and Models.}  In the prediction problem we consider here, the outcome is the action that is chosen by the row player in a given instance of play, and the features are the 18 entries of the payoff matrix. A prediction rule is thus any map $f: \mathbb{R}^{18} \rightarrow \{a_1,a_2,a_3\}$ from $3 \times 3$ payoff matrices to row player actions.

 For each prediction rule $f$ and test set of observations $\{(g_i,a_i\}_{i=1}^n$\textemdash where $g_i$ is the payoff matrix in observation $i$, and $a_i$ is the observed row player action\textemdash we evaluate error using the \emph{misclassification rate} 
\begin{equation*}
\frac1n \sum_{i=1}^n \mathbbm{1}\left( f(g_i) \neq a_i)\right).
\end{equation*}
This is the fraction of observations where the predicted action was not the observed action.

As a naive baseline, we consider guessing uniformly at random for all games, which yields an expected misclassification rate of $2/3$.  Additionally, we consider a prediction rule based on the \emph{Poisson Cognitive Hierarchy Model}  (PCHM), which supposes that there is a distribution over players of differing levels of sophistication: The \emph{level-0} player is maximally unsophisticated and randomizes uniformly over his available actions, while the \emph{level-1} player best responds to level-0 play \citep{StahlWilson94,StahlWilson95,Nagel}. \citet{CamererHoChong04} defines the play of level-$k$ players, $k\geq2$, to be best responses to a perceived distribution 
\begin{equation}
p_{k}(h)=\frac{\pi_{\tau}(h)}{\sum_{l=0}^{k-1}\pi_{\tau}(l)}\qquad\forall\,\,h\in\mathbb{N}_{<k} \label{perceiveddistr}
\end{equation}
over (lower) opponent levels, where $\pi_{\tau}$ is the Poisson distribution
with rate parameter $\tau$.\footnote{Throughout, we take
$\tau$ to be a free parameter and estimate it from the training data.} A predicted distribution over actions
is derived by supposing that the proportion of level-$k$
players in the population is proportional to $\pi_{\tau}(k)$. Assuming this is the true distribution of play, the misclassification rate is minimized by predicting the mode of this distribution, and this is what we set as the PCHM prediction. 

As in Section \ref{sec:CV}, we estimate the free parameter $\tau$ on training data, and evaluate the out-of-sample prediction of the estimated model. All reported prediction errors are tenfold cross-validated.

\paragraph{Results.} 

Because we use the classification loss as the loss function, the best attainable classification error  will differ across games:  In games where all subjects choose the same action, the perfect $0$-error prediction is feasible, but when play is close to uniform over the actions, it will be hard to improve over random guessing. This means that the same level of predictive accuracy should potentially be evaluated quite differently, depending on what kinds of games are being predicted.

We illustrate this by comparing predictions for two subsets of our data:
\emph{Data Set A} consists of the 16,660 observations of play from the 359 games with no strictly dominated actions.\footnote{Specifically, we consider games where no pure action is strictly dominated by another pure action.} \emph{Data Set B} consists of the 7,860 observations of play from the 161 games in which the action profile with the highest sum of player payoffs is outside of the support of level-$k$ actions,\footnote{Here we use the classic definition from \citet{StahlWilson95} and \citet{Nagel}, where each level-$k$ action is the best response to the level-$(k-1)$ action.} and moreover the difference in the payoff sums is large (at least 20\% of the largest row player payoff in the game.) For example, the following game is included in Data Set B:
\[\begin{array}{cccc}
& a_1 & a_2 & a_3 \\
a_1 & 40,40 & 10,20 & 70,30 \\
a_2 & 20,10 & 80,80 & 0,100 \\
a_3 & 30,70 & 100,0 & 60,60 
\end{array}\]
In this game, action $a_3$ is level 1, since it yields the highest expected payoff against uniform play, and action $a_1$ is level 2, since it is a best response against play of $a_1$. Because $(a_1,a_1)$ is a pure-strategy Nash equilibrium, action $a_1$ is then level-$k$ for all $k\geq 2$. The highest possible player sum achieved by playing either $a_1$ or $a_3$ is 120 (from action profile $(a_3,a_3)$), but the action profile $(a_2,a_2)$ yields a higher payoff sum of 160. The difference, 40, is $1/2$ of the max row player payoff in this game, 80.

In both data sets, a range of values for the free parameter $\tau$ generate  the same predicted modal action, and so have the same cross-validated prediction error. For all of the games in our data, this mode is simply the level-1 action.  But as Table \ref{tab:games} shows,  PCHM improves upon the naive benchmark by a larger amount for prediction of play in Data Set B, compared to Data Set A. Using perfect prediction as the benchmark, this would imply that PCHM is a more complete model of play for games in Data Set B.\footnote{Instead of our task of predicting each action, \citet{FudenbergLiang} studies the task of predicting the modal action in each game; the ideal prediction for that task always has no error at all. Correspondingly for that prediction task, \citet{FudenbergLiang} also used a different cross-validation procedure: Instead of dividing the data into folds at random as described above, it split the set of games so that the games in the training set were not used for testing.This alternative is relevant for the study of how well we can extrapolate from one game to another, which is not the question of interest here.}

\begin{table}[H]
\begin{center}
\begin{tabular}{lcc}
\toprule
 & Data Set A & Data Set B \\
\midrule
Naive Benchmark & 0.66 & 0.66 \\
PCHM & 0.49 & 0.44  \\
& (0.004) & (0.009) \\
\bottomrule
\end{tabular} 
\end{center}
\caption{PCHM improves upon the naive baseline by a larger amount for prediction of play in Data Set B.}
 \label{tab:games}
\end{table}

But the amount of irreducible error in the two data sets may be quite different, leading to different predictive limits. Thus we need to understand how the prediction errors compare to the \emph{best achievable} error for the two data sets. We can again gain insight into this by building a lookup table. The rows of the table are the different games, and the associated predictions are the modal actions (observed for those games) in the training data. Given sufficiently many observations, the modal action in the training data will also be the action most likely to be played in the test data, thus minimizing classification error.

Below we report the Table Lookup performance and completeness measures relative to this performance.

\begin{table}[H]
\begin{center}
\begin{tabular}{lcccc}
\toprule
&\multicolumn{2}{c}{Data Set A} & \multicolumn{2}{c}{Data Set B} \\
 & Error & Completeness & Error & Completeness\\
\midrule
Naive Benchmark & 0.66 & 0\% & 0.66 & 0\%\\
PCHM & 0.49 & 68\% & 0.44 & 67\% \\
& (0.006) & & (0.009) \\

Table Lookup & 0.41 & 100\% & 0.34 & 100\%\\
& (0.005) & & (0.006)  \\
\bottomrule
\end{tabular} 
\end{center}
\caption{PCHM achieves roughly the same completeness for both data sets.} \label{tab:resultsGames}
\end{table}

Although PCHM achieves a smaller absolute improvement over the naive baseline for Data Set A, the achievable improvement is also lower. Thus, relative to the appropriate benchmarks, the completeness of PCHM is in fact roughly equivalent for the two data sets (and marginally lower for Data Set B). This comparison illustrates how prediction accuracy can be misleading without an accompanying benchmark. 

Our exercise here is not special to the two sets of games we have examined; indeed, we can repeat the analysis for other subsets of the data, and determine completeness measures for each of these. For example, Table \ref{tab:LargeDiff} reports prediction errors for data set consisting of the 9,243 observations of play from the 175 games where the level 1 action's expected payoff against uniform play is much higher than the expected payoff of the next best action (specifically, it is larger by at least $1/4$ of the max row player payoff in the game).

\begin{table}[H]
\begin{center}
\begin{tabular}{lcc}
\toprule
 & Error & Completeness \\
\midrule
Naive Benchmark & 0.66 & 0\%\\
PCHM & 0.28 & 97\%  \\
& (0.004) & \\

Table Lookup & 0.27 & 100\% \\
& (0.005) &  \\
\bottomrule
\end{tabular} 
\end{center}
\caption{Prediction errors for games in which the level-1 action is much better against uniform play than the next best action.} \label{tab:LargeDiff}
\end{table}

The Table Lookup error is much lower for this set of games, revealing that for these games, play is much more concentrated on a single action. Thus we would hope for our models to also achieve higher predictive accuracies, and indeed we find that PCHM predicts an incorrect action only 28\% of the time. For a more exhaustive inquiry into when PCHM succeeds and fails, we could elicit completeness measures for different subsets of the data, and identify those games where PCHM is most incomplete.

\subsection{Domain \#3: Human Generation of Random Sequences}  \label{sec:coins}

\paragraph{Background and Data.} Extensive experimental and empirical evidence suggests that humans misperceive
randomness, expecting for example that sequences of coin flips ``self-correct" (too many Heads in a row must be followed by a Tails) and are balanced (the proportion of Heads and Tails are approximately the same) \citep{barhillel,LSN}. These misperceptions are significant not only for their basic psychological interest,
but also for the ways in which misperception of randomness manifests
itself in a variety of contexts: for example, 
investors' judgment of sequences of (random) stock returns \citep{shleifer}, professional
decision-makers' reluctance to choose the same (correct) option multiple times
in succession \citep{Shue},
 and people's execution of a mixed strategy in a game \citep{RPS}. 
 
A common experimental framework in this area is to ask human participants
to generate fixed-length strings of $k$ (pseudo-)random coin flips, 
for some small value
of $k$ (e.g. $k = 8$), and then to compare the produced distribution over length-$k$
strings to the output of a Bernoulli process that generates
realizations from $\{H,T\}$ independently and uniformly at random \citep{budescu,distributional}. Following in this tradition, we use the platform Mechanical Turk to collect a large dataset of
human-generated strings designed to simulate the output of a
{\em Bernoulli(0.5) process}, in which each symbol in the string
is generated from $\{H,T\}$ independently and uniformly at random. To incentive effort, we told subjects that payment would be approved only if their (set of) strings could not be identified as human-generated with high confidence.\footnote{In one experiment, 537
subjects each whom produced 50 binary strings of length eight. In a second experiment, an
additional 101 subjects were asked to each generate 25 binary strings
of length eight.}$^,$\footnote{Subjects were informed: ``To encourage effort in this task, we have developed an algorithm (based on previous Mechanical Turkers) that detects human-generated coin flips from computer-generated coin flips. You are approved for payment only if our computer is not able to identify your flips as human-generated with high confidence."}  Following removal of subjects who were clearly not attempting to mimic a random process, our final data set consisted of 21,975 strings generated by 167 subjects.\footnote{Our initial data set consists of 29,375 binary strings. We chose to remove all subjects who repeated any string in more than five rounds. This cutoff was selected by looking at how often each subject generated any given string, and finding the average ``highest frequency" across subjects. This turned out to be 10\% of the strings, or five strings. Thus, our selection criteria removes all subjects whose highest frequency was above average. This selection eliminated 167 subjects and 7,400 strings, yielding a final dataset with 471 subjects and 21,975 strings. We check that our main results are not too sensitive to this selection criteria by considering two alternative choices in Appendix \ref{appDiffCut}\textemdash first, keeping only the initial 25 strings generated by all subjects, and then, removing the subjects whose strings are ``most different" from a Bernoulli process under a $\chi^2$-test. We find very similar results under these alternative criteria.}

\paragraph{Prediction Task, Performance Metric,  and Models.} 
We consider the problem of predicting the probability that the eighth entry in a string is $H$ given its first seven elements.  Thus the outcome here is a number in $[0,1]$\textemdash that is a distribution on $\{H,T\}$\textemdash and  the feature space is $\{H,T\}^7$ (note that as in the previous examples we fit a representative-agent model and do not treat the identity of the subject as feature). 

Given a test dataset $\{(s_i^1, \dots, s_i^8)\}_{i=1}^n$ of $n$ binary strings of length-8, we evaluate the error of the prediction rule $f$ using mean-squared error
\begin{equation*}
\frac1n \sum_{i=1}^n \left( s_i^8 - f(s_i^1, \dots, s_i^7)\right)^2
\end{equation*}
where $f(s_i^1, \dots, s_i^7)$ is the predicted probability that the eighth flip is `$H$' given the observed initial seven flips $s_i^1, \dots, s_i^7$, and $s_i^8$ is the actual eighth flip.\footnote{Alternatively we could have defined the outcome to be an individual realization of $H$ or $T$,  so that prediction rules are maps $f:\{H,T\}^7 \rightarrow \{H,T\}$,  and then evaluated error using the  misclassification rate (i.e. the fraction of instances where the predicted outcome was not the realized outcome). We do not take a stand on which method is better, but note that the completeness measure can depend on which one is used. 
In Appendix \ref{appDiffLoss} we show that the completeness measures are very similar  using this alternative formulation.} Note that the naive baseline of unconditionally guessing 0.5 guarantees a mean-squared prediction error of 0.25. Moreover, if the strings in the test set were truly generated via a Bernoulli(0.5) process, then no prediction rule could improve in expectation upon the naive error.\footnote{Due to the convexity of the loss function, it is possible to do \emph{worse} than the naive baseline, for example by predicting 1 unconditionally.} We expect that the presence of behavioral errors in the generation process will make it possible to improve upon the naive baseline, but do not know \emph{how much} it is possible to improve upon 0.25.

In this task, the natural naive baseline is the rule that  unconditionally guesses that the probability the final flip is `$H$' is 0.5. We compare this to prediction rules based on \citet{Rabin} and \citet{gambler}, both of which predict generation of negatively autocorrelated sequences.\footnote{Although both of these
frameworks are models of mistaken \emph{inference} from data, as opposed to human attempts to  generate random sequences, they are
easily adapted to our setting, as the papers explained.} Our prediction rule based on \citet{Rabin} supposes that subjects generate sequences by drawing sequentially \emph{without replacement} from an urn containing $0.5 N$ `1' balls and $0.5 N$ `0' balls. The urn is ``refreshed" (meaning the composition is returned to its original) every period with independent probability $p$. This model has two free parameters: $N \in \mathbb{Z}_+$ and $p \in [0,1]$. 

Our prediction rule based on \citet{gambler} assumes that the first flip $s_1 \sim \mbox{Bernoulli(0.5)}$ while each subsequent flip $s_k$ is distributed
\[s_k \sim \mbox{Ber}\left(0.5 - \alpha \sum_{t=0}^{k-2} \delta^t (2 \cdot s_{k-t-1}-1)\right),\]
where the parameter $\delta \in \mathbb{R}_+$ reflects the (decaying) influence of past flips, and the parameter $\alpha\in \mathbb{R}_+$ measures the strength of negative autocorrelation.\footnote{We make a small modification on the \citeasnoun{gambler} model, allowing $\alpha, \delta \in \mathbb{R}_+$ instead of $\alpha, \delta \in [0,1)$.}


\paragraph{Results.}

Table \ref{tab:seq} shows that both prediction rules improve upon the naive baseline. The need for a benchmark for achievable prediction is starkest in this application, however, as the best improvement is only 0.0008, while the gap between the best prediction error and a perfect zero is large. This is not surprising, as we expect substantial variation in the eighth flip following the same initial seven flips because we asked subjects to mimic a fair coin.

\begin{table}[H]%
	\bigskip
	\begin{minipage}{\columnwidth}
\begin{center}
\begin{tabular}{ccc}	
\toprule
	& Error \\
	\midrule
	Naive Benchmark & 0.25  \\
Rabin (2002) &0.2494 \\[-1mm]
&\footnotesize{(0.0007)}\\
Rabin and Vayanos (2010) & 0.2492\\[-1mm]
&\footnotesize{(0.0007)}\\
\bottomrule
\end{tabular} 
\end{center}
\caption{Both models improve upon naive guessing, but the absolute improvement is small.} \label{tab:seq}
		\centering
	\end{minipage}
\end{table}%

For this problem, the lookup table's rows correspond to the $2^7$ unique initial seven-flip sequences, and we associate each such string to the empirical frequency with which that string is followed by `$H$' in the training data. Given a sufficiently large training set, we can approximate the true continuation frequency for each initial sequence, and hence approximate the best achievable error. We note here that although there are $2^7$ unique initial sequences, with approximately 21,000 strings in our data set, we have (on average) 164 observations per initial sequence.

\begin{table}[H]%
	\label{tab:comparison}
	\begin{minipage}{\columnwidth}
\begin{center}
\bigskip
\begin{tabular}{ccccc}	
\toprule
	& Error & Completeness \\
	\midrule
	Naive Benchmark & 0.25 & 0  \\
Rabin (2002) &0.2494 & 10\% \\[-1mm]
&\footnotesize{(0.0007)}& \\
Rabin \& Vayanos (2010) & 0.2492& 14\%\\[-1mm]
&\footnotesize{(0.0007)}& \\
Table Lookup &0.2441 & 100\% \\[-1mm]
&\footnotesize{(0.0006)} &\\
\bottomrule
\end{tabular} 
\end{center}
\caption{The Table Lookup benchmark permits a more accurate representation of the completeness of these models.} 	\label{tab:resultsCoin}
		\centering
	\end{minipage}
\end{table}%

We find that Table Lookup achieves a prediction error of 0.2439, so that naively comparing achieved prediction error against perfect prediction (which would suggest a completeness measure of at most 0.4\%) grossly misrepresents the performance of the models. Relative to the Table Lookup benchmark, the existing models produce up to 14\% of the achievable improvement in prediction error. This suggests that although negative autocorrelation is indeed present in the human-generated strings, and explains a sizable part of the deviation from a Bernoulli(0.5) process, there is additional structure that could yet be exploited for prediction.

\section{Extensions}

\subsection{Subject Heterogeneity}\label{subsection heterogeneity}

So far we've focused on evaluating representative agent models that implement a single prediction across all subjects. When we evaluate models that include subject heterogeneity, the question of what is the best achievable level of accuracy is still relevant, and the suitable analogue of Table Lookup\textemdash with subject type added as an additional feature\textemdash can again help us to determine this. The exact implementation of Table Lookup will depend on how the groups are determined. As a simple illustration, we return to our first domain\textemdash evaluation of risk\textemdash and demonstrate how to construct a predictive bound for certain models with subject heterogeneity. 

The models that we consider extend the Expected Utility and Cumulative Prospect Theory models introduced in Section \ref{sec:risk} by allowing for three groups of subjects. To test the models, we randomly select 71 (out of 171) subjects to be test subjects, and 45 (out of 50) lotteries to be test lotteries. All other data\textemdash the 100 training subject's choices in all lotteries, as well as the test subject's choices in the 5 training lotteries\textemdash are used for training the models. 

In more detail, we first use the training subjects' responses in the \emph{training} lotteries to develop a clustering algorithm for separating subjects into three groups.\footnote{We use a simple algorithm, $k$-means, which minimizes the Euclidean distance between the vectors of reported certainty equivalents for subjects within the same group.} This algorithm can be used to assign a group number to any new subject based on their choices in the five training lotteries.  Second, we use each group's training subjects' responses in the \emph{test} lotteries to estimate free model parameters\textemdash that is, the single free parameter of the EU model, and the four free parameters for CPT. This yields three versions of EU and CPT, one per group.

Out of sample, we first use the clustering algorithm to assign groups to the test subjects, and then use the associated models to predict their certainty equivalents in the test lotteries. We measure accuracy using mean-squared error, as in Section \ref{sec:risk}, and we again report the Expected Value prediction as a naive baseline. 

\begin{table}[H]
\begin{center}
\begin{tabular}{lc}
\toprule
 & Prediction Error  \\
\midrule
Naive Benchmark & 91.13 \\
& (10.44) \\
Expected Utility & 86.68\\
& (10.69)\\
CPT & 57.14\\
& (7.17) \\
\bottomrule
\end{tabular}
\caption{Prediction Errors Achieved by Models with Subject Heterogeneity} \label{tab:subjheterog}
\end{center}
\end{table}

What we find from Table \ref{tab:subjheterog} is very similar to what we observed in Section \ref{sec:risk}: Both models improve upon the naive baseline, but we do not know how complete these improvements are. To better evaluate the achieved improvements, we need a benchmark that tells us the best feasible prediction.

Our approach here for constructing an upper bound is to learn the mean response of  training subjects in each group for each lottery, and predict those means. With sufficiently many training subjects, this method approximates the best possible accuracy. We find that although the CPT error is substantially different from zero, the model is again nearly complete. 


\begin{table}[H]
\begin{center}
\begin{tabular}{lcc}
\toprule
 & Prediction Error & Completeness \\
\midrule
Naive Benchmark & 104.17& 0\%\\
& (12.95) & \\
Expected Utility & 86.68 & 36\% \\
& (10.69)\\
CPT & 57.14 & 96\%\\
& (7.17) \\
Table Lookup & 55.45 & 100\%  \\
& (6.26)\\
\bottomrule
\end{tabular} 
\end{center}
\end{table}

Because the same clustering method is used in all of the approaches, the gap between Table Lookup and the existing models does not shed light on how much  predictions could be improved by better ways of  grouping the subjects. The comparison of Table Lookup's performance here, 55.45, with its performance from Section \ref{sec:risk}, 65.58, sheds light on the size of predictive gains achieved by the present method for clustering.

\subsection{Comparing Feature Sets} \label{sec:featureset}

In addition to evaluating the  predictive limits of a feature set and the completeness of existing models,  Table Lookup can be used to compare the predictive power of different feature sets.   We illustrate this potential comparison by revisiting our problem from Section \ref{sec:coins}\textemdash predicting human generation of randomness\textemdash and evaluating the predictive value of certain features. To do this, we consider ``compressed" Table Lookup algorithms built on different properties of the string, where strings of the same type are bucketed into the same row, and focus on the the predictive value of two properties: number of Heads, and flips 4-8. Our compressed Table Lookup based on the number of Heads partitions the set of length-7 strings depending on the total number `$H$' flips in the string, and learns a prediction for each partition element; similarly, our compressed Table Lookup based on flips 4-7 partitions the set of strings depending only on outcomes including and after flip 4. Just as our original Table Lookup algorithm returned an approximation of the highest level of predictive accuracy using the full structure of initial flip data, these compressed Table Lookup algorithms approximate the highest level of predictive accuracy that is achievable using a particular kind of structure in the strings.

\begin{table}[H]%
	\caption{\small{Comparison of the value of various feature sets.}}
	\label{tab:featureset}
	\begin{minipage}{\columnwidth}
\begin{center}
\bigskip
\begin{tabular}{ccccc}	
\toprule
	& Error & Completeness  \\
	\midrule
	Naive Benchmark & 0.25 & 0\%  \\
	Flips 4-7 &0.2478 & 36\%  \\[-1mm]
	&\footnotesize{(0.0010)} &  \\
	Number of Heads &0.2464 & 59\% &   \\[-1mm]
	&\footnotesize{(0.0009)} &  \\

Full Table Lookup &0.2441 & 100\%  \\[-1mm]
&\footnotesize{(0.0006)} &  \\
\bottomrule
\end{tabular} 
\end{center}
		\centering
	\end{minipage}
\end{table}%

We find that these simple features achieve large fractions of the achievable improvement over the naive rule of always predicting that the probability of $H$ is $1/2$. For example, using only  the number of Heads as a feature achieves 59\% of the improvement of full Table Lookup. Using only the most recent three flips  achieves 36\% of the predictive improvement that is achieved by using all seven initial flips; the fact that this improvement is not complete demonstrates that there is predictive content in flips 1-3 beyond what is captured in flips 4-7.

The feature set corresponding to our ``Full Table Lookup" is itself partial relative to an even richer feature set. It is interesting to consider what might constitute a set of unmeasured features of the human participant's behavior that would significantly improve predictive accuracy, for example the speed at which the strings were entered. The exercise in Section \ref{subsection heterogeneity}, in which we used subject types (determined based on choices in auxiliary problems), constitutes yet another way to expand the table. As we have shown above, the application and comparison of Table Lookup for different feature sets is one potentially useful approach for measuring the predictive value of those features.\footnote{Note that the value of individual features will in general depend on what other features are available.}

\section{Conclusion}
When evaluating the predictive performance of a theory, it is important to know not just whether the theory is predictive, but also how complete its predictive performance is. We propose the use of Table Lookup as a way to measure  the best \emph{achievable} predictive performance for a given problem, and the completeness of a model as a measure of how close it comes to this bound. We demonstrate three domains in which completeness can help us to evaluate the performance of existing models.

The present paper has focused on the criterion of predictiveness. When we take other criteria into account, such as the interpretability or generality of the model, then we may prefer models that are not 100\% complete by the measure proposed here\textemdash for example, we may prefer to sacrifice some predictive power in return for higher explainability, as in \citet{FudenbergLiang}. 

Finally, we note that all the tests mentioned so far involve training and testing models on data drawn from the same domain. A question for future work would be how to compare the transferability of models across domains. Indeed, we may expect that economic models that are outperformed by machine learning models in a given domain  have higher transfer performance outside of the domain. In this sense, within-domain completeness may provide an incomplete measure of the ``overall completeness" of the model, and we leave development of such notions to future work.

\bibliographystyle{ecta}

\clearpage

\appendix 

\begin{center}
\huge{Appendix}
\end{center}

\section{Is Table Lookup the most predictive algorithm for our data?} \label{app:TLerrBound}

In the main text, we use the performance of Table Lookup as an approximation of the best possible accuracy. Below we investigate whether the data sets we study are  large enough for this to be a good approximation. 

We first review some results from the machine learning and statistics literatures, which explain how the  cross-validated standard errors that we report in the main text can be used as a measure for how well the Table Lookup error approximates the irreducible error (Section \ref{app:CVSE}). 

In Section \ref{app:RF}, we compare Table Lookup's performance with that of bagged decision trees, an algorithm that scales better to smaller quantities of data. We find that in each of our prediction problems,  the two prediction errors are similar, and Table Lookup weakly outperforms bagged decision trees. Finally, in Section \ref{app:subsample}, we study the sensitivity of the Table Lookup performance to the quantity of data. The predictive accuracies achieved using our full data sets are very close to those achieved using, for example, just 70\% of the data. This again suggests that only minimal improvements in predictive accuracy are feasible from further increases in data size.

\subsection{Cross-Validated Standard Error}
\label{app:CVSE}

Suppose that the loss function is mean-squared error: $\mathcal{L}(y',y)=(y'-y)^2$. (Similar arguments apply for the  misclassification rate; see e.g. \citet{Domingos}.) Let 
\[f_{P}^*(x) = \mathbb{E}_{P}[y \mid x]\]
 be the ideal prediction rule discussed in Section \ref{subsec:AccComplete}, which assigns to each $x$ its expected outcome $y$ under distribution $P$.
 Write $f_{TL}[Z]$ for the \emph{random} Table Lookup prediction rule that has been estimated from a set $Z$ of $n$ i.i.d. training observations.  The expected mean-squared error of $f_{TL}$ on a new observation $(x,y)\sim P$ can be decomposed as follows \citep{Hastie}: 
\begin{align*}
     & \mathbb{E}[(f_{TL}[Z](x)-y)^2] = \\
& \quad \quad 
\underbrace{\mathbb{E}[(f^\ast(x)-y)^2]}_{\text{irreducible noise}} + \underbrace{\left(\mathbb{E}[f_{TL}[Z](x)]-f^\ast(x)\right)^2}_{\text{bias}} + \underbrace{\mathbb{E}[(f_{TL}[Z](x)-\mathbb{E}[f_{TL}[Z](x)])^2]}_{\text{sampling error}}
\end{align*}
where the expectation is both over the realization of the training data $Z$ used to train Table Lookup, and also over the realization of the test observation $(x,y)$.

The first component is the \emph{irreducible noise} introduced in (\ref{disp:IrrErr}). The second component, \emph{bias}, is the  mean-squared difference between the \emph{expected} Table Lookup prediction and the prediction of the ideal prediction rule $f^\ast.$ The final component, \emph{sampling error} or \emph{variance}, is the variance of the Table Lookup prediction (reflecting the sensitivity of the algorithm to the training data).

Since Table Lookup is an unbiased estimator, the second component is zero. Thus, irreducible noise is 
the difference between the expected Table Lookup error and the sampling error of the Table Lookup predictor. As described in Section \ref{subsec:finite}, we follow standard procedures of using the \emph{cross-validated prediction error} to estimate the expected Table Lookup error, and using the \emph{variance of the cross-validated prediction errors} to estimate the sampling error \citep{Hastie}. That is,
\[\mathbb{E}[(f_{TL}[Z](x)-\mathbb{E}[f_{TL}[Z](x)])^2] \approx \frac{1}{K} \mbox{Var}(\{CV_1,\dots, CV_K\})\]
where $CV_i$ is the prediction error for the $i$-th iteration of cross-validation. The right-hand side of the display is the square of the cross-validated standard errors reported in the main text; thus, we have from Tables \ref{tab:resultsCE}, \ref{tab:resultsGames}, and \ref{tab:resultsCoin}:

\begin{table}[H]
\begin{center}
\begin{tabular}{lcc}
\toprule
 & Table Lookup Error & Sampling Error\\
\midrule
Risk Preferences & 65.58 & 9 \\
Predicting Initial Play, Data Set A & 0.41 & $<$0.0001 \\
Predicting Initial Play, Data Set B & 0.34 & $<$0.0001 \\
Human Generation of Random Sequences & 0.2441 & $<$0.0001 \\
\bottomrule
\end{tabular} 
\end{center}
\end{table}

\subsection{Comparison with  Scalable Machine Learning Algorithms}
\label{app:RF}

Another way to evaluate whether our Table Lookup algorithm approximates the best possible prediction accuracy is to compare it with the performance of other machine learning algorithms. Below we compare its performance with \textbf{bagged decision trees} (also known as  \emph{bootstrap-aggregated} decision trees). This algorithm creates several bootstrapped data sets from the training data by sampling with replacement, and then trains a \textbf{decision tree} on each bootstrapped training set. Decision trees are nonlinear prediction models that recursively partition
the feature space and learn a (best) constant prediction for each
partition element. The prediction of the bagged decision tree algorithm is an aggregation of the predictions of individual decision trees. When the loss function is mean-squared error, the decision tree ensemble predicts the average of the predictions of the individual trees. When the loss function is misclassification rate, the decision tree ensemble predicts based on a majority vote across the ensemble of trees.

Table \ref{tab:compareDT} shows that for each prediction problem, the error of the bagged decision tree algorithm is comparable to and slightly worse than that of the Table Lookup algorithm. These results again suggest that the Table Lookup error is a reasonable approximation for the best achievable error.

\begin{table}[H]
\begin{center}
\begin{tabular}{lcccc}
\toprule
 & Risk & Games A &  Games B &  Sequences \\
\midrule
Bagged Decision Trees & 65.65 & 0.45 & 0.36 & 0.2442 \\
    & (0.10) & (0.004) & (0.005) & (0.0005) \\
Table Lookup & 65.58 & 0.41 & 0.34 & 0.2441\\
& (3.00) & (0.005) & (0.006) & (0.0006)\\
\bottomrule
\end{tabular} 
\end{center}
\caption{Table Lookup outperforms Bagged Decision Trees in each of our prediction problems.} \label{tab:compareDT}
\end{table}

\subsection{Performance of Table Lookup on Smaller Samples} \label{app:subsample}

Finally, we  report the Table Lookup cross-validated performance on random samples of $x$\% of our data, where $x \in \{10,20,\dots,100\}$. For each $x$, we repeat the procedure 1000 times, and report the average performance across iterations.  We find that the Table Lookup performance flattens out for larger values of $x$, suggesting that the quantity of data we have is indeed large enough that further increases in the data size will not substantially improve predictive performance.

\begin{table}[H]
\begin{center}
\begin{tabular}{lcccc}
\toprule
 $x$\% & Risk & Games A &  Games B &  Sequences \\
\midrule
10\% & 69.47  & 0.4191 & 0.3473 & 0.2592 \\
    & (11.13) & (0.012) & (0.018) & (0.0034) \\
20\% & 67.13 & 0.4183 & 0.3476 &   0.2504 \\
    & (7.95) & (0.0018) & (0.024) & (0.0018) \\
30\% & 66.28 & 0.4178 & 0.3472 & 0.2479 \\
    & (6.51) & (0.0022) & (0.0029) & (0.0014) \\
40\% & 66.25 & 0.4169 & 0.3470 & 0.2464 \\
    & (5.65) & (0.0024) & (0.0032) & (0.0011) \\
50\% & 65.68 & 0.4157 & 0.3459 & 0.2458 \\
    & (4.59) & (0.0025) & (0.0036) & (0.0010) \\
60\% & 65.68 & 0.4141 & 0.3449 & 0.2452 \\
    & (4.24) & (0.0027) & (0.0040) & (0.0008) \\
70\% & 65.68 & 0.4131 & 0.3435 & 0.2448 \\
    & (3.95) & (0.0031) & (0.0045) & (0.0007) \\
80\% & 65.68 & 0.4119 & 0.3427 & 0.2445 \\
    & (3.95) & (0.0034) & (0.0046) & (0.0007) \\
90\% & 65.66 & 0.4109 & 0.3416 & 0.2443 \\
    & (3.71) & (0.0034) & (0.0047) & (0.0007) \\
100\% & 65.58 & 0.4100 & 0.3404 & 0.2441 \\
    & (3.00) & (0.0036) & (0.0051) & (0.0006) \\
\bottomrule
\end{tabular} 
\end{center}
\caption{Performance of Table Lookup using $x$\% of the data, averaged over 100 iterations for each $x$}
\label{tab:subsample}
\end{table}

\section{Experimental Instructions for Section \ref{sec:coins}} \label{instructions}
Subjects on Mechanical Turk were presented with the following introduction screen:
\begin{figure}[H]
\begin{center}
\label{heads}
\includegraphics[scale=0.4]{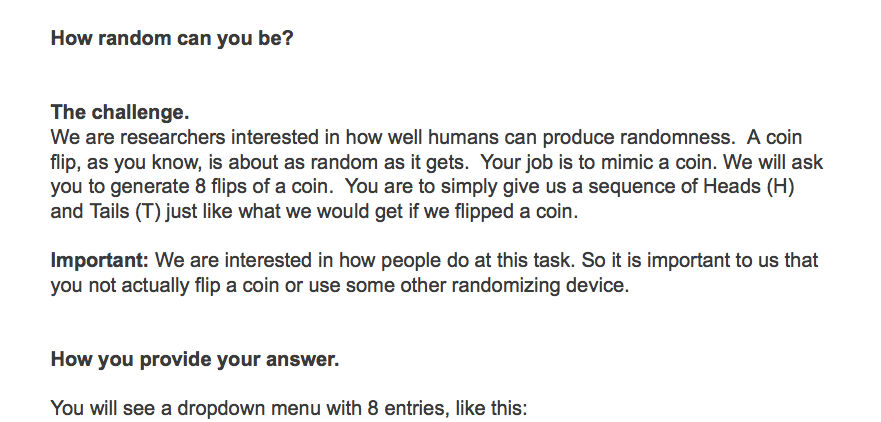}
\includegraphics[scale=0.4]{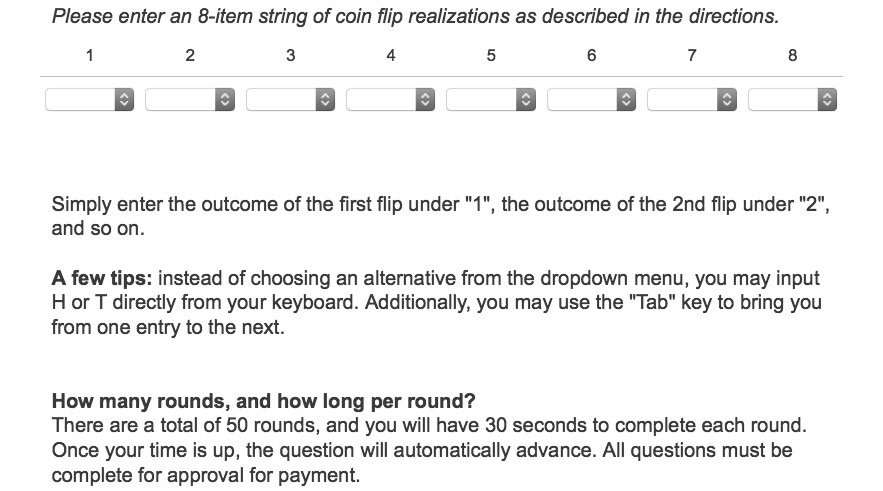}
\includegraphics[scale=0.4]{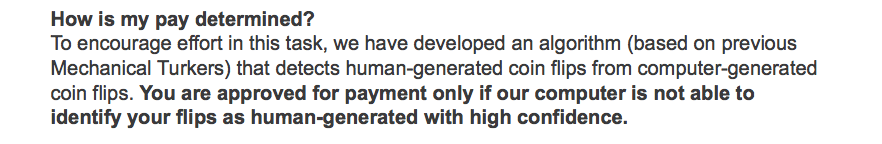}
\end{center}
\end{figure}

\section{Supplementary Material to Section \ref{sec:coins}}

\subsection{Robustness} \label{appDiffLoss}

Here we check how our results in Section \ref{sec:coins} change when the outcome space and error function are changed so that prediction functions are maps $f: \{H,T\}^7 \rightarrow \{H,T\}$ and the error for predicting the test data set $\{(s_i^1, \dots, s_i^8\}_{i=1}^n$ is defined to be
\[\frac1n \sum_{i=1}^n \mathbbm{1}(s_i^8 \neq f(s_i^1, \dots, s_i^7)),\]
i.e. the misclassification rate. We use as a naive benchmark the prediction rule that guesses $H$ and $T$ uniformly at random; this is guaranteed an expected misclassification rate of 0.50.

For this problem, the Table Lookup benchmark learns the modal continuation for each sequence in $\{0,1\}^7$. We find that the completeness of \citet{Rabin} and \citet{Rabin2000} relative to the Table Lookup benchmark are respectively 19\% and 9\%. 

\begin{table}[H]%
	\label{tab:comparison}
	\begin{minipage}{\columnwidth}
\begin{center}
\bigskip
\begin{tabular}{ccccc}	
\toprule
	& Error & Completeness \\
	\midrule
	Naive Benchmark & 0.50 & 0  \\
Rabin (2002) &0.45 & 19\% \\[-1mm]
&\footnotesize{(0.003)}& \\
Rabin \& Vayanos (2010) & 0.475 &9\%\\[-1mm]
&\footnotesize{(0.01)}& \\
Table Lookup & 0.23 &1 \\[-1mm]
&\footnotesize{(0.002)} &\\
\bottomrule
\end{tabular} 
\end{center}
		\centering
	\end{minipage}
\end{table}%

\subsection{Different Cuts of the Data} \label{appDiffCut}

\paragraph{Initial strings only.} We repeat the analysis in Section \ref{sec:coins} using data from all subjects, but only their first 25 strings. This selection accounts for potential fatigue in generation of the final strings, and leaves a total of 638 subjects and 15,950 strings. Prediction results for our main exercise are shown below using this alternative selection.

\begin{table}[H]
	\begin{minipage}{\columnwidth}
\begin{center}
\bigskip
\begin{tabular}{ccccc}	
\toprule
	& Error & Completeness \\
\midrule
	Naive Benchmark & 0.25 & 0 &  \\
Rabin \& Vayanos (2010) & 0.2491 & 5\%  \\[-1mm]
& (0.0008) &\\
Table Lookup & 0.2326 & 100\% &\\[-1mm]
& (0.0030) &  \\
\bottomrule
\end{tabular} 
\end{center}
		\centering
	\end{minipage}
\end{table}%

\paragraph{Removing the least random subjects.} For each subject, we conduct a Chi-squared test for the null hypothesis that their strings were generated under a Bernoulli process. We order subjects by $p$-values and remove the 100 subjects with the lowest $p$-values (subjects whose generated strings were most different from what we would expect under a Bernoulli process). This leaves a total of 538 subjects and 24,550 strings. Prediction results for our main exercise are shown below using this alternative selection.

\begin{table}[H]
	\begin{minipage}{\columnwidth}
\begin{center}
\bigskip
\begin{tabular}{ccccc}	
\toprule
	& Error & Completeness  \\
\midrule
	Naive Benchmark & 0.25 & 0 &  \\
Rabin \& Vayanos (2010) & 0.2491 & 12\% & \\[-1mm]
& (0.0005)& \\
Table Lookup & 0.2427 & 100\% & \\[-1mm]
& (0.0016) &\\
\bottomrule
\end{tabular} 
\end{center}
	\end{minipage}
\end{table}%

\end{document}